\documentclass[12pt]{article}
\usepackage{epsfig}
\usepackage{amsmath,amsfonts,amssymb}

\newcommand{\be}{\begin{equation}}
\newcommand{\ee}{\end{equation}}
\newcommand{\bem}{\begin{displaymath}}
\newcommand{\eem}{\end{displaymath}}
\newcommand{\ba}{\begin{eqnarray}}
\newcommand{\se}{\setcounter{equation}{0}}
\newcommand{\ea}{\end{eqnarray}}
\newcommand{\re}[1]{(\ref{#1})}

\newcommand{\1}{^{-1}}

\newcommand{\delt}{\delta^{\hspace*{-0.2mm}\mbox{\tiny G}}}
\newcommand{\dg}{^{\dagger}}
\newcommand{\di}{\mbox{d}\,}
\newcommand{\e}{\mbox{e}}

\newcommand{\f}{{\mbox{\scriptsize f}}} 
\newcommand{\g}{{\mbox{\scriptsize g}}} 
 
\newcommand{\bG}{\bar{G}} 
\newcommand{\G}{{\cal B}} 
\newcommand{\ga}{\gamma_5}
\newcommand{\h}{\frac{1}{2}}
\newcommand{\Id}{\mbox{1\hspace{-1.05mm}l}}

\newcommand{\bP}{\bar{P}} 

\newcommand{\s}{{\cal S}} 
\newcommand{\bs}{\bar{{\cal S}}} 
\newcommand{\ts}{\tilde{{\cal S}}} 
\newcommand{\bS}{\bar{S}} 
\newcommand{\T}{{\cal T}} 
\newcommand{\Tr}{\mbox{Tr}} 
\newcommand{\bu}{\bar{u}} 
\newcommand{\U}{{\cal U}}

\newcommand{\bw}{\bar{w}}

\newcommand{\SG}{S_{\rm p}} 
\newcommand{\SH}{\tilde{S}_{\rm p}} 
\newcommand{\bSG}{\bar{S}_{\rm p}} 

\begin{document}
 
\hfill {\sc HU-EP}-04/67
 
\vspace*{1cm}
 
\begin{center}
 
{\Large \bf Gauge transformations in lattice chiral theories}
 
\vspace*{0.9cm}
 
{\bf Werner Kerler}
 
\vspace*{0.3cm}
 
{\sl Institut f\"ur Physik, Humboldt-Universit\"at, D-12489 Berlin,
Germany}
 
\end{center}

\vspace*{1cm} 

\begin{abstract}
We show that gauge-transformation properties of correlation functions in
chiral gauge theories on the finite lattice are determined in a general way. 
\end{abstract}

\vspace*{0.3cm}

\section{Introduction}

Gauge invariance of the chiral determinant on the lattice has been considered 
a major problem \cite{go01}. A particular construction aiming at gauge 
invariance has been presented in Ref.~\cite{lu98}. Using finite 
transformations we here show that gauge-transformation properties of 
correlation functions in chiral gauge theories on the finite lattice are 
determined in a general way, so that there is no freedom for adjustments by
constructions. A careful consideration of equivalence classes of pairs of 
bases is important in this context. Our results hold also in the presence 
of zero modes and for any value of the index. 

We also add a corresponding analysis of the subject in terms of gauge 
variations. Within this we find that fully exploiting the covariance 
requirement for the current of Ref.~\cite{lu98} leads to the same result as 
follows from the indicated consideration of equivalence classes of pairs of 
bases. On the other hand, the behavior of the effective action anticipated 
in Ref.~\cite{lu98} in its special case and the view of the anomaly 
cancelation there turn out not to conform with the actual results.

In Section 2 we collect some relations needed. In Section 3 we use finite 
transformations to analyze the general cases with both and with only one 
of the chiral projections depending on the gauge field. In Section 4 we 
consider variations of the effective action and the special case of 
Ref.~\cite{lu98}. Section 5 contains conclusions and discussions.

\section{General relations}\se

\subsection{Chiral projections}

The chiral projections $\bP_+$ and $P_-$ obey $\bP_+\dg=\bP_+=
\bP_+^2$, $P_-\dg=P_-=P_-^2$ and
\be
\bP_+D=DP_-,
\label{DP}
\ee
where $D$ is the Dirac operator. 
For the numbers of anti-Weyl and Weyl degrees of freedom $\bar{N}=\Tr\,\bP_+$
and $N=\Tr\,P_-$ we have the two possibilities $\bar{N}=d$, $N=d-I$ or 
$\bar{N}=d+I$, $N=d$, where $d=\h\Tr\,\Id$. Requiring $D$ to be 
$\ga$-Hermitian and normal, $I$ is its index. 

The chiral projections may also be expressed as 
\be
P_-=\h(\Id-\ga G),\quad\qquad\bP_+=\h(\Id+\bG\ga),
\label{GaG}
\ee
with $\ga$-Hermitian and unitary operators $G$ and $\bG$,
which according to \re{DP} satisfy
\be
D+\bG D\dg G=0.
\label{DG}
\ee

\subsection{Basic fermionic functions}

In terms of Grassmann variables basic fermionic correlation 
functions -- which do not vanish identically and of which linear combinations 
make up general functions -- for the Weyl degrees of freedom are of form
\ba
\langle\chi_{i_{r+1}}\ldots\chi_{i_N}\bar{\chi}_{j_{r+1}}\ldots
\bar{\chi}_{j_{\bar{N}}}\rangle_{\f}=\hspace*{88mm}\nonumber\\
s_r\int\di\bar{\chi}_{\bar{N}} \ldots\di\bar{\chi}_1\di\chi_N
\ldots\di\chi_1\;\;\e^{-\bar{\chi}M\chi}\;\;\chi_{i_{r+1}}\ldots\chi_{i_N}
\bar{\chi}_{j_{r+1}} \ldots\bar{\chi}_{j_{\bar{N}}},\hspace*{6mm}
\ea
where we put $s_r=(-1)^{rN-r(r+1)/2}$. 
The fermion fields $\bar{\psi}_{\sigma'}$ and $\psi_{\sigma}$  
are given by $\bar{\psi}=\bar{\chi}\bu\dg$ and $\psi=u\chi$ with bases 
$\bu_{\sigma'j}$ and $u_{\sigma i}$ which satisfy
\be
P_-=uu\dg,\quad u\dg u=\Id_{\rm w},\qquad\qquad\bP_+=\bu\bu\dg,\quad\bu\dg\bu
=\Id_{\rm\bw},
\label{uu}
\ee 
where $\Id_{\rm w}$ and $\Id_{\rm \bw}$ are the identity operators in the
spaces of the Weyl and anti-Weyl degrees of freedom, respectively. Specifying
the fermion action as $\bar{\chi}M\chi=\bar{\psi}D\psi$ we then have for 
basic correlation functions of the fermion fields
\be
\langle\psi_{\sigma_{r+1}}\ldots\psi_{\sigma_N}\bar{\psi}_{\bar{\sigma}_{r+1}}
\ldots\bar{\psi}_{\bar{\sigma}_{\bar{N}}}\rangle_{\f}
=\frac{1}{r!}\sum_{\bar{\sigma}_1\ldots\bar{\sigma}_r}\sum_{\sigma_1,\ldots,
\sigma_r}\bar{\Upsilon}_{\bar{\sigma}_1\ldots\bar{\sigma}_{\bar{N}}}^*
\Upsilon_{\sigma_1\ldots\sigma_N}D_{\bar{\sigma}_1\sigma_1}\ldots
D_{\bar{\sigma}_r\sigma_r}
\label{COR}
\ee
with the alternating multilinear forms
\ba
\Upsilon_{\sigma_1\ldots\sigma_N}=\sum_{i_1,\ldots,i_N=1}^N\epsilon_{i_1, 
\ldots,i_N}u_{\sigma_{1}i_{1}}\ldots u_{\sigma_Ni_N},\\\bar{\Upsilon}_{
\bar{\sigma}_1\ldots{\bar{\sigma}_{\bar{N}}}}=\sum_{j_1,\ldots,
j_{\bar{N}}=1}^{\bar{N}}\epsilon_{j_1,\ldots,j_{\bar{N}}}\bar{u}_{\bar{
\sigma}_{1}j_{1}}\ldots\bar{u}_{\bar{\sigma}_{\bar{N}}j_{\bar{N}}}.
\label{FO}
\ea

\subsection{Subsets of bases}

By \re{uu} the bases are only fixed up to unitary transformations, 
$u^{[S]}=uS$, $\bu^{[\bar{S}]}=\bu\bar{S}$. While the chiral
projections are invariant under such transformations, the forms 
$\Upsilon_{\sigma_1\ldots\sigma_N}$ and $\bar{\Upsilon}_{\bar{\sigma}_1
\ldots{\bar{\sigma}_{\bar{N}}}}$ get multiplied by factors $\det_{\rm w}S$ 
and $\det_{\rm \bw}\bar{S}$, respectively. Therefore in order that general
correlation functions remain invariant, we have to impose
\be
{\det}_{\rm w}S\cdot{\det}_{\rm\bw}\bar{S}\dg=1.
\label{UNI}
\ee
This is so since firstly in full correlation functions only a phase factor
independent of the gauge field can be tolerated. Secondly this factor must 
be $1$ in order that in general functions, which involve linear combinations 
of basic functions, individual basis transformations in its parts leave the 
interference terms in the moduli of the amplitudes invariant. 

Condition \re{UNI} has important consequences. Without it all bases related 
to a chiral projection are connected by unitary transformations. With it the 
total set of pairs of bases $u$ and $\bu$ decomposes into inequivalent 
subsets, beyond which legitimate transformations do not connect. These subsets 
of pairs of bases obviously are equivalence classes. Clearly the formulation 
of the theory must be restricted to one of such classes (which raises the 
question which choice is appropriate for describing physics). 

Different ones of the indicated equivalence classes are obviously related by 
pairs of basis transformations $S$, $\bar{S}$ for which
\be
{\det}_{\rm w}S\cdot{\det}_{\rm\bw}\bar{S}\dg=\e^{i\Theta}\qquad
\mbox{ with }\qquad\Theta\ne0
\label{NUNI}
\ee
holds.
The phase factor $\e^{i\Theta}$ then determines how the results of the 
formulation of the theory with one class differ from the results of the 
formulation with the other class.

\section{Gauge transformations}\se

A gauge transformation $D'=\T D\T\dg$ of the Dirac operator by \re{DP}
implies the corresponding transformations 
\be
P_-'=\T P_-\T\dg,\qquad \bP_+'=\T\bP_+\T\dg
\label{PT}
\ee 
of the chiral projections. In view of \re{DG} (and since on the lattice 
$D\dg\ne-D$) at least one of them should depend on the gauge field.
Thus the cases are of interest where none or where only one of the chiral
projections commutes with $\T$.

\subsection{Non-constant chiral projections}

We first consider the case where $[\T,P_-]\ne0$ and $[\T,\bP_+]\ne0$.
To get the behavior of the bases we start from the fact that conditions 
\re{uu} must be satisfied such that relations \re{PT} hold. It is 
obvious that given a solution $u$ of the conditions \re{uu}, then $T u$ 
is a solution of the transformed conditions \re{uu}. Analogous considerations
apply to $\bu$. All solutions are then obtained by performing basis 
transformations.

In addition \re{UNI} is to be satisfied, i.e.~these considerations are 
to be restricted to an equivalence class of pairs of bases. Accordingly 
the original class $uS$, $\bu\bS$ and the transformed one $u'S'$, $\bu'\bS'$ 
are related by
\be
u'S'=\T uS\s,\qquad\bu'\bS'=\T\bu\bS\bs,
\label{uS}
\ee
where $u$, $\bu$, $S$, $\bS$ satisfy \re{uu} and \re{UNI}, respectively, and 
$u'$, $\bu'$, $S'$, $\bS'$ their transformed versions. For full generality
of \re{uS} we have included the unitary transformations $\s(\T,\U)$ and 
$\bs(\T,\U)$ obeying
\be
{\det}_{\rm w}\s(\Id,\U)\big({\det}_{\rm\bw}\bs(\Id,\U)\big)^*=1,
\ee 
with $\s$ combining as $\s(\T_{\rm a},\U)\s(\T_{\rm b},\T_{\rm a}\U\T_{\rm a}
\dg)=\s(\T_{\rm b}\T_{\rm a},\T_{\rm b}\T_{\rm a}\U \T_{\rm a}\dg\T_{\rm b}
\dg)$ and $\bs$ analogously.

Inserting \re{uS} into \re{COR} we get for the transformation of correlation 
functions
\ba
\langle\psi_{\sigma_1'}'\ldots\psi_{\sigma_R'}'\bar{\psi}_{\bar{\sigma}_1'}'
\ldots\bar{\psi}_{\bar{\sigma}_{\bar{R}}'}'\rangle_{\f}'=\hspace{86mm}
\nonumber\\
\e^{i\vartheta}\sum_{\sigma_1,\ldots,\sigma_R}\sum_{\bar{\sigma}_1,
\ldots,\bar{\sigma}_{ \bar{R}}}\T_{\sigma_1'\sigma_1}\ldots\T_{\sigma_R'
\sigma_R} \langle\psi_{\sigma_1}\ldots\psi_{\sigma_R}
\bar{\psi}_{\bar{\sigma}_1} \ldots\bar{\psi}_{\bar{\sigma}_{\bar{R}}}
\rangle_{\f}\,\T_{\bar{\sigma}_1\bar{\sigma}_1'}\dg\ldots
\T_{\bar{\sigma}_{\bar{R}}\bar{\sigma}_{\bar{R}}'}\dg,
\label{COV}
\ea
where
\be
\e^{i\vartheta}={\det}_{\rm w}\s\cdot{\det}_{\rm\bw}\bs\dg.
\label{NUN}
\ee
In \re{NUN} so far $\vartheta\ne0$ for $\T\ne1$ is admitted with the
ambiguity of the many possible choices of $\s$ and $\bs$. However, 
\re{NUN} with $\vartheta\ne0$ obviously is just of form \re{NUNI} 
related to transformations to arbitrary inequivalent subsets of 
pairs of bases, which ultimately cannot be tolerated. According to \re{UNI} 
then fixing to $\vartheta=0$ is appropriate and the correlation 
functions turn out to transform gauge-covariantly.

\subsection{One constant chiral projection}

In the special case where $[\T,P_-]\ne0$ and $[\T,\bP_+]=0$, the 
equivalence class of {\it pairs} of bases always contains members where 
$\bP_+$ is represented as $\bP_+=\bu_{\rm c}\bu_{\rm c}\dg$ with constant 
$\bu_{\rm c}$. Indeed, given a pair $u$, $\bu$ we note that $\bu$ is 
generally related to $\bu_{\rm c}$ by a basis transformation 
$\bu=\bu_{\rm c}\bS_{\rm e}$. Thus transforming $u$ as 
$u=u_{\rm e}S_{\rm e}$, where $S_{\rm e}$ is subject to 
${\det}_{\rm w}S_{\rm e}\cdot{\det}_{\rm\bw}\bS_{\rm e}\dg=1$, 
according to \re{UNI} the pair $u_{\rm e}$, $\bu_{\rm c}$ is in the same 
equivalence class as the pair $u$, $\bu$. For a transformed pair 
$u'$, $\bu'$ we analogously get the equivalent pair 
$u_{\rm e}'$, $\bu_{\rm c}$. Then instead of \re{uS} we have
\be
u_{\rm e}'S'=\T u_{\rm e}S\ts,\qquad \bu_{\rm c}\bS_{\rm c}={\rm const},
\label{uS1}
\ee
where $S$ and $\bS_{\rm c}$ as well as $S'$ and $\bS_{\rm c}$ satisfy 
\re{UNI}, so that
\be
{\det}_{\rm w}S'={\det}_{\rm w}S.
\label{DD}
\ee 
For full generality in \re{uS1} the unitary transformation $\ts(\T,\U)$ is 
included which obeys
\be
{\det}_{\rm w}\ts(\Id,\U)=1.
\ee 

We next note that because of $[\T,\bP_+]=0$ we can rewrite $\bu_{\rm c}$ as
\be
\bu_{\rm c}=\T\bu_{\rm c}S_{\T}
\label{uS2}
\ee
where $S_{\T}$ is unitary. With this and \re{uS1} we get for the 
transformation of the correlation functions \re{COR} the form 
\ba
\langle\psi_{\sigma_1'}'\ldots\psi_{\sigma_R'}'\bar{\psi}_{\bar{\sigma}_1'}'
\ldots\bar{\psi}_{\bar{\sigma}_{\bar{R}}'}'\rangle_{\f}'=\hspace{86mm}
\nonumber\\
\e^{i\vartheta}\;{\det}_{\rm\bw}S_{\T}\dg
\sum_{\sigma_1,\ldots,\sigma_R}\sum_{\bar{\sigma}_1,
\ldots,\bar{\sigma}_{ \bar{R}}}\T_{\sigma_1'\sigma_1}\ldots\T_{\sigma_R'
\sigma_R} \langle\psi_{\sigma_1}\ldots\psi_{\sigma_R}
\bar{\psi}_{\bar{\sigma}_1} \ldots\bar{\psi}_{\bar{\sigma}_{\bar{R}}}
\rangle_{\f}\,\T_{\bar{\sigma}_1\bar{\sigma}_1'}\dg\ldots
\T_{\bar{\sigma}_{\bar{R}}\bar{\sigma}_{\bar{R}}'}\dg,
\label{COV1}
\ea
where $\e^{i\vartheta}={\det}_{\rm w}\ts$. The ambiguity of the many
possible choices of $\ts$ here is fixed by noting that $\vartheta\ne0$
is again related to transformations to arbitrary inequivalent subsets of 
pairs of bases, which are prevented by choosing $\vartheta=0$.

For the calculation of the factor 
${\det}_{\rm\bw}S_{\T}\dg={\det}_{\rm\bw}(\bu_{\rm c}\dg\T\bu_{\rm c})$
in \re{COV1} we note that with $[\T,\bP_+]=0$ and 
$\T=e^{\G}$ we get $\bu_{\rm c}\dg\T\bu_{\rm c}=\bu_{\rm c}\dg\e^{\G\bP_+}
\bu_{\rm c}$ and the simultaneous eigenequations $\G\bP_+\bu_j^{\rm d}=
\omega_j\bu_j^{\rm d}$ and $\bP_+\bu_j^{\rm d}=\bu_j^{\rm d}$. Since 
$\bu^{\rm d}=\bu_{\rm c}\hat{S}$ with unitary $\hat{S}$, we obtain 
${\det}_{\rm\bw}(\bu_{\rm c}\dg\e^{\G\bP_+}\bu_{\rm c})=\prod_j\e^{\omega_j}
=\exp(\Tr(\G\bP_+))$, so that using $P_+=\h(1+\ga)\Id$ we have
\be
{\det}_{\rm\bw}S_{\T}\dg=\exp({\textstyle\h}\Tr\,\G),
\label{RF}
\ee 
where in detail $\Tr\,\G=4i\sum_{n,\ell}b_n^{\ell}\,\mbox{tr}_{\g}
T^{\ell}$ with constants $b_n^{\ell}$ and group generators $T^{\ell}$.

\section{Variational approach}\se

\subsection{General relations}

We define general gauge-field variations of a function $\phi(\U)$ by
\be
\delta\phi(\U)=\frac{\di\phi\big(\U(t)\big)}{\di t}\bigg|_{t=0}\,,\qquad 
\U_{\mu}(t)=\e^{t\G_{\mu}^{\rm left}}\U_{\mu}\e^{-t\G_{\mu}^{\rm right}},
\label{DEF}
\ee
where $(\U_{\mu})_{n'n}=U_{\mu n}\delta^4_{n',n+\hat{\mu}}$ and 
$(\G_{\mu}^{\rm left/right})_{n'n}=B_{\mu n}^{\rm left/right}\delta^4_{n',n}$.
The special case of gauge transformations is then described by
\be
\G_{\mu}^{\rm left}=\G_{\mu}^{\rm right} =\G.
\ee

Varying the logarithm of the general condition \re{UNI} gives
\be
\Tr_{\rm w}(S\dg\delta S)-\Tr_{\rm \bw}(\bar{S}\dg\delta\bar{S})=0.
\label{vUNI}
\ee
Instead of ${\det}_{\rm w}S\cdot{\det}_{\rm\bw}\bar{S}\dg=1$, as needed in 
general functions involving linear combinations of basic functions, this 
obviously reflects the weaker condition ${\det}_{\rm w}S\cdot{\det}_{\rm\bw}
\bar{S}\dg=$ const. Relation \re{vUNI} can also be expressed in terms of bases as
\be
\Tr\big(\delta(uS)(uS)\dg\big)-\Tr\big(\delta(\bu\bS)(\bu\bS)\dg\big)
=\Tr(\delta u\,u\dg)-\Tr(\delta\bu\,\bu\dg),
\label{uUNI}
\ee 
which indicates that $\Tr(\delta u\,u\dg)-\Tr(\delta\bu\,\bu\dg)$ remains
invariant within the extended subset of bases specified
by ${\det}_{\rm w}S\cdot{\det}_{\rm\bw}\bar{S}\dg=$ const.

Requiring absence of zero modes of $D$ (and thus also restricting to the 
vacuum sector) the effective action can be considered, for the variation of 
which one gets
\be
\delta\ln{\det}_{\rm\bw w}M=\Tr(P_-D\1\delta D)+\Tr(\delta u\,u\dg)-
\Tr(\delta\bu\,\bu\dg).
\label{EFF}
\ee

\subsection{Gauge transformations}
 
In the special case of gauge transformations we can use the definition 
\re{DEF} and the finite transformation relations to get the related
variations. For operators with ${\cal O}\big(\U(t)\big)=\T(t)\,{\cal O}
\big(\U(0)\big)\,\T\dg(t)$ and $\T(t)=\e^{t\G}$ this gives 
\be
\delt{\cal O}=[\G,{\cal O}].
\ee

In the case $[\T,P_-]\ne0$, $[\T,\bP_+]\ne0$ according to \re{uS} we have 
for the bases $u(t)=\T(t)u(0)\SG(t)$, $\bu(t)=\T(t)\bu(0)\bSG(t)$ 
where $\SG=S\s S'\hspace{0mm}\dg$, $\bSG=\bS\bs\bS'\hspace{0mm}\dg$ and 
obtain
\be
\delt u=\G\,u+u\,\SG\dg\,\delt\SG,\qquad
\delt\bu=\G\,\bu+\bu\,\bSG\dg\,\delt\bSG.
\label{DU}
\ee
Using these relations the terms in the variation of the effective action 
become 
\be
\Tr(P_-D\1\delt D)=\Tr(\G\bP_+)-\Tr(\G P_-),
\label{DE1}
\ee
\be
\Tr(\delt u\,u\dg)=\Tr(\G P_-)+\Tr_{\rm w}(\SG\dg\,\delt\SG),\quad
\Tr(\delt\bu\,\bu\dg)=\Tr(\G\bP_+)+\Tr_{\rm\bw}(\bSG\dg\,\delt\bSG),
\ee
so that with \re{vUNI} we get
$\delt\ln{\det}_{\rm\bw w}M=\Tr_{\rm w}(\s\dg\,\delt\s)-
\Tr_{\rm\bw}(\bs\dg\,\delt\bs)$.
Because the exclusion of transformations to inequivalent subsets of pairs 
of bases leads to the relation $\Tr_{\rm w}(\s\dg\,\delt\s) -
\Tr_{\rm\bw}(\bs\dg\,\delt\bs)=0$, we thus obtain
\be
\delt\ln{\det}_{\rm\bw w}M=0,
\ee
as expected according to the results for finite transformations. 

In the case $[\T,P_-]\ne0$, $[\T,\bP_+]=0$ according to \re{uS1} we get 
instead of 
\re{DU}
\be
\delt u_{\rm e}=\G\,u_{\rm e}+u_{\rm e}\,\SH\dg\,\delt\SH,\qquad
\delt\bu_{\rm c}=0,
\label{DU1}
\ee
where $\SH=S\ts S'\hspace{0mm}\dg$, which with \re{DD}
and $P_+=\h(1+\ga)\Id$ gives for the effective action
$\delt\ln{\det}_{\rm\bw w}M=\Tr_{\rm w}(\ts\dg\,\delt \ts)+\h\Tr\,\G.$
Excluding transformations to inequivalent subsets of pairs of 
bases here means that $\Tr_{\rm w}(\ts\dg\,\delt \ts)=0$, so that we remain 
with 
\be
\delt\ln{\det}_{\rm\bw w}M=\h\Tr\,\G,
\label{DEF0}
\ee
again as expected according to the results for finite transformations.

\subsection{Special case of L\"uscher}

L\"uscher \cite{lu98} considers the variation of the effective action, 
imposing the Ginsparg-Wilson relation \cite{gi82} $\{\ga,D\}=D\ga D$
and using chiral projections which correspond to the choice $\bG=\Id$ and 
$G=\Id-D$ in \re{GaG}. He assumes $\bu=$ const, so that the last term 
in \re{EFF} is absent and condition \re{vUNI} reduces to 
$\Tr(S\dg\delta S)=0$. He defines a current $j_{\mu n}$ by
\be
\Tr(\delta u\,u\dg)=-i\sum_{\mu,n}\mbox{tr}_{\g}(\eta_{\mu n}j_{\mu n}),
\qquad\delta U_{\mu n}=\eta_{\mu n}U_{\mu n},
\ee
and requires it to transform gauge-covariantly.

His generator is given by $\eta_{\mu n}=B_{\mu,n+\hat{\mu}}^{\rm left}-
U_{\mu n}B_{\mu n}^{\rm right}U_{\mu n}\dg$ in terms of our left and right
generators. We get explicitly
\be
j_{\mu n}=i(U_{\mu n}\rho_{\mu n}+\rho_{\mu n}\dg U_{\mu n}\dg),
\qquad\rho_{\mu n,\alpha'\alpha}=\sum_{j,\sigma}u_{j\sigma}\dg
\frac{\partial u_{\sigma j}\hspace*{7mm}}{\partial U_{\mu n,\alpha\alpha'}}.
\ee
The requirement of gauge-covariance 
$j_{\mu n}'=\e^{B_{n+\hat{\mu}}}j_{\mu n}\e^{-B_{n+\hat{\mu}}}$ because 
of $U_{\mu n}'=\e^{B_{n+\hat{\mu}}}U_{\mu n}\e^{-B_n}$ implies that one 
must have 
\be
\rho_{\mu n}'=\e^{B_n}\rho_{\mu n}\e^{-B_{n+\hat{\mu}}},
\ee
which with $u'=\T uS\ts S'\hspace{0mm}\dg$ and \re{DD} leads to the condition
\be
\sum_{j,k}\ts_{kj}\dg\frac{\partial\ts_{jk}\hspace*{7mm}}{\partial 
U_{\mu n,\alpha\alpha'}}=0.
\label{II}
\ee
\hspace{0mm}From \re{II} and $\ts\1=\ts\dg$ it follows that
\be
\Tr_{\rm w}(\ts\dg\delt\ts)=0.
\label{DELT}
\ee
With this we arrive at the interesting result that the covariance requirement 
for L\"uscher's current leads just to what we have found before to follow 
from the exclusion of transformations to inequivalent subsets of pairs of 
bases.

We now have seen in different ways that in the special case considered by 
L\"uscher one obtains the definite result 
\be
\delt\ln{\det}_{\rm\bw w}M=\h\Tr\,\G
\label{DTD}
\ee
which leaves no room for adjustments by particular constructions. 
Furthermore, it has also become obvious that aiming at 
$\delt\ln{\det}_{\rm\bw w}M=0$ in Ref.~\cite{lu98} does not to conform 
with the actual result \re{DTD}.

\section{Conclusions and discussions}\se

We have given an unambiguous derivation of the gauge-transformation 
properties of correlation functions in chiral gauge theories on the finite
lattice using finite transformations. In the
case where both of the chiral projections are gauge-field dependent the 
exclusion of switching to arbitrary inequivalent subsets of pairs of bases 
leads to gauge covariance. In the cases where one of the chiral projections 
is constant a factor depending on the particular gauge transformation remains.
A careful consideration of equivalence classes of pairs of bases has been
important in our analysis. Our results have been seen to hold also in the 
presence of zero modes and for any value of the index.

We have also considered the subject in terms of variations of 
the effective action (which implies restriction to absence of zero modes
and the vacuum sector). In this context we have shown that satisfying the 
covariance requirement for L\"uscher's current quite remarkably leads to the 
same result as the exclusion of switching to inequivalent subsets of pairs
of bases. It has furthermore turned out that the behavior anticipated 
in Ref.~\cite{lu98} for the effective action in the special case there does 
not conform with the actual result.

Altogether it has become obvious that (whatever the detailed gauge-field
dependences might be) the gauge-transformation properties of correlation 
functions in chiral gauge theories on the finite lattice are determined 
in a general way and cannot be adjusted by particular constructions, as 
has been tried to do in literature. 

In Ref.~\cite{lu98} a main argument was that without the anomaly cancelation
condition one would be unable to cancel the anomaly term. However, as has
been seen in Section 4.2 the anomaly term $\Tr(P_-D\1\delt D)=
\Tr(\G\bP_+)-\Tr(\G P_-)$ in the case where both chiral projections are 
gauge-field dependent is compensated by the basis contribution 
$\Tr(\G P_-) -\Tr(\G\bP_+)$, while with $\bP_+$ being constant 
by the basis contribution $\Tr(\G P_-)$ there is compensation up to the 
quantity $\h\Tr\,\G$. 

There is no contradiction to continuum perturbation theory since in the 
limit one arrives just at the usual situation where the anomaly cancelation 
condition is needed to get gauge invariance of the chiral determinant. This
is seen from the consideration of perturbation theory in Ref.~\cite{ke03},
of which we here briefly mention some main features. In the limit the chiral 
projections become constant and their products with propagators get the 
appropriate forms. Using the notation $D=D_0+D_{\rm I}$, $u=u_0+u_{\rm I}$
and $\bu=\bu_0+\bu_{\rm I}$, where the quantities with indices 0 refer to 
the free case, the vertices decompose as 
\be
\bP_{+0}D_{\rm I}P_{-0}+\bu_0\bu_{\rm I}\dg Du_{\rm I}u_0\dg+\bu_0
\bu_{\rm I}\dg DP_{-0}+\bP_{+0}Du_{\rm I}u_0\dg.
\label{Mpe}
\ee
Since the chiral projections get constant in the limit and accordingly then
are described by constant bases, the terms in \re{Mpe} which rely on 
$u_{\rm I}$ and $\bu_{\rm I}$ vanish in the limit. For the surviving 
contributions thus agreement with continuum perturbation theory 
becomes obvious at lower order. With appropriate locality properties of 
the Dirac operator this extends to higher orders.

With respect to the chiral determinant thus what happens is that in the 
limit the contributions are no longer there which when performing gauge 
transformations on the finite lattice produce the compensating terms. 
Furthermore, in the limit obviously also the particular cases with one 
constant chiral projection are no longer distinct from the other ones. 
In this way in all cases one arrives at the usual continuum result.

\section*{Acknowledgement}

I wish to thank Michael M\"uller-Preussker and his group for their kind
hospitality.


\begin{thebibliography}{0}

\bibitem{go01} For an overview see: M. Golterman, Nucl. Phys. B (Proc. Suppl.) 
               94 (2001) 189.
\bibitem{lu98}  M. L\"uscher,
              Nucl. Phys. B549 (1999) 295; 
              Nucl. Phys. B568 (2000) 162. 
\bibitem{gi82}  P.H. Ginsparg, K.G. Wilson, 
              Phys. Rev. D25 (1982) 2649. 
\bibitem{ke03}  W. Kerler, Nucl. Phys. B680 (2004) 51. 
\end{thebibliography}
\end{document}